\documentclass[11pt,a4paper]{article}
\usepackage[utf8]{inputenc}
 
\usepackage{caption}
\usepackage{subcaption}
\usepackage{jcappub}
\usepackage{bm}
\usepackage{epsfig}
\usepackage{bbold}
\usepackage{graphicx}
\usepackage{amsmath}
\usepackage{amssymb}
\usepackage{amsbsy}
\usepackage{soul}
\usepackage{slashed}
\usepackage{afterpage}
\usepackage{psfrag}
\usepackage[noabbrev]{cleveref}
\usepackage{dsfont}
\usepackage{enumerate}
\usepackage[shortlabels]{enumitem}
\usepackage[utf8]{inputenc}
\usepackage{amsmath}
\usepackage[dvipsnames]{xcolor}
\usepackage{graphicx}
\usepackage{braket}
\usepackage{siunitx}
\usepackage{float}
\usepackage{cite}

\graphicspath{ {./Images/} }

\title{\begin{center}
    Thermal curvature perturbations \\in thermal inflation
\end{center}}

\author[a]{Mar Bastero-Gil,}
\author[b]{Joaquim M.~Gomes,}
\author[c]{and Jo\~ao G. Rosa}

\affiliation[a]{Departamento de F\'{\i}sica Te\'orica y del Cosmos,
  Universidad de Granada,\\ Granada-18071, Spain}
\affiliation[b]{Department of Mathematical Sciences, University of Liverpool,\\ Liverpool L69 7ZL, United Kingdom}
\affiliation[c]{Univ Coimbra, Faculdade de Ci\^encias e Tecnologia da Universidade de Coimbra and CFisUC,\\ Rua Larga, 3004-516 Coimbra, Portugal}

\emailAdd{mbg@ugr.es}
\emailAdd{j.m.gomes@liverpool.ac.uk}
\emailAdd{jgrosa@uc.pt}

\makeatletter
\gdef\@fpheader{}
\makeatother

\abstract{We compute the power spectrum of super-horizon curvature perturbations generated during a late period of thermal inflation, taking into account fluctuation-dissipation effects resulting from the scalar flaton field's interactions with the ambient radiation bath. We find that, at the onset of thermal inflation, the flaton field may reach an equilibrium with the radiation bath even for relatively small coupling constants, maintaining a spectrum of thermal fluctuations until the critical temperature $T_c$, below which thermal effects stop holding the field at the false potential minimum. This enhances the field variance compared to purely quantum fluctuations, therefore increasing the average energy density during thermal inflation and damping the induced curvature perturbations. In particular, we find that this inhibits the later formation of primordial black holes, at least on scales that leave the horizon for $T>T_c$. The larger thermal field variance also reduces the duration of a period of fast-roll inflation below $T_c$, as the field rolls to the true potential minimum, which should also affect the generation of (large) curvature perturbations on even smaller scales.}

\begin{document}

\maketitle

\section{Introduction}\label{sec:intro}

It is widely believed that the universe went through a period of inflation in its early stages \cite{Starobinsky:new_type, Sato:first_order_phase, Guth:Inflation_Seminal_Guth, Linde:New_Inflation}, thus explaining its observed homogeneity and isotropy on large scales, as well as its apparently small spatial curvature. Most importantly, inflation in principle provided the seeds for the small curvature perturbations that grew into the large-scale structure that we observe in the Universe.

Although the simplest models postulate a single period of slow-roll inflation lasting for at least 50-60 e-folds after the largest presently observable scales became super-horizon, there is {\it a priori} no reason to exclude scenarios with multiple inflation periods with different dynamics. In particular, it is well known that reheating after inflation may lead to the production of e.g. topological defects if the associated reheating temperature exceeds the grand unification scale ($\sim 10^{16}$ GeV) \cite{Rocher:2004et} or other unwanted relics such as moduli or gravitinos in supersymmetric (SUSY) models \cite{Pagels:1981ke, Weinberg:1982zq, Khlopov:1984pf}. Such relics could have overclosed the Universe or spoiled the successful predictions of primordial nucleosynthesis through their late decay \cite{Kawasaki:2008qe}. This and the fact that currently there is no evidence for such relics motivates considering scenarios with additional inflationary stages that could have diluted their abundances \cite{Randall:Solving_moduli,Lyth:Gut_Higgs,Lyth:Thermal_Inflation,Asaka:moduli_problem_thermal_inflation,Barreiro:aspects_Thermal_Inflation}.

One of the most appealing possibilities is a late period of thermal inflation, where a scalar {\it flaton }field is trapped in a false vacuum by thermal effects above a certain critical temperature. Candidates to drive such a secondary inflation period are ubiquitous in SUSY and supergravity theories, in particular given the many flat directions in the scalar potential that characterize such models at the renormalizable level \cite{Gherghetta:1995dv}. The spectrum of curvature perturbations generated during such a period (or possibly multiple periods) need not be nearly as scale-invariant as the one generated by the first period of slow-roll inflation, during which the large-scale perturbations observable in the Cosmic Microwave Background (CMB) anisotropies became super-horizon. In fact, this spectrum was recently computed in \cite{Dimopoulos:thermal_inflation_primordial_black_holes}, where it was shown that large curvature perturbations could have been generated (on small scales) during a period of thermal inflation and a {\it fast roll} inflation period \cite{Linde:2001ae} that potentially followed it once thermal effects stopped trapping the field in the false vacuum state. These large curvature/density perturbations could have then collapsed into a significant population of primordial black holes upon horizon-reentry later in the radiation-dominated epoch. Such a possibility has attracted a substantial interest in the recent literature given the latter's appeal as dark matter candidates and the possibility that these may explain the recent LIGO/Virgo detections of heavy black hole binaries (see e.g.~\cite{Carr:2021bzv}).

The analysis in \cite{Dimopoulos:thermal_inflation_primordial_black_holes} considered, however, only the part of the curvature spectrum generated by quantum fluctuations of the flaton scalar field. Since thermal effects are a crucial aspect in the dynamics of thermal inflation, one should investigate whether thermal fluctuations also play an important role. This is our goal with this work. We note, in particular, that the flaton field is trapped in a false vacuum at temperatures above a certain critical temperature, as we review in the next section, due to the large thermal mass resulting from its interactions with the ambient thermal bath.  It is well-known that such interactions also lead to fluctuation-dissipation effects, resulting in an effective Langevin-like equation describing the dynamics of the scalar field. Such effects have been thoroughly analyzed in the context of warm inflation scenarios \cite{ Berera:1995ie, Berera:1995wh, Berera:1998gx, Berera:1998px, Berera:Warm_Inflation_adiabaticity, Berera:2001gs, Berera:Warm_Microphysical, Bastero-Gil:2009sdq, Bastero-Gil:2010dgy, Bastero-Gil:2011zxb, Bastero-Gil:2011clw, Bastero-Gil:2012akf, Bartrum:2013fia, Bastero-Gil:2014jsa, Bastero-Gil:2016qru, Rosa:2018iff, Bastero-Gil:2019gao, Berghaus:2019whh}, in setting initial conditions for slow-roll inflation in a pre-inflationary radiation epoch \cite{Bastero-Gil:Initial_Conditions_Inflation}, and in cosmological phase transitions both after and during (warm) inflation \cite{Bartrum:2014fla, Rosa:2021gbe}. Our objective is then to apply the techniques developed in these contexts to the case of thermal inflation, and investigate their role in the generation of curvature perturbations during this period.

Surprisingly, we find that for thermal flaton fluctuations the amplitude of the curvature power spectrum is suppressed with respect to the purely quantum case analyzed in \cite{Dimopoulos:thermal_inflation_primordial_black_holes}, at least for scales exiting the horizon before the temperature decreases below the critical value. This is essentially due to the fact that, as we will show, thermal effects, by enhancing flaton density fluctuations, also increase the time-dependent part of the average energy density during thermal inflation. This effect overcomes the enhancement of individual perturbation modes, therefore suppressing the corresponding power spectrum.

This work is organized as follows. We will start by constructing a generic model for thermal inflation in Section \ref{sec:thermal_inflation}. The curvature perturbation spectrum induced by the thermal flaton fluctuations is computed in Section \ref{sec:curvature}. In Section \ref{sec:comparison} we compare our result with the purely quantum computation performed in \cite{Dimopoulos:thermal_inflation_primordial_black_holes}, discussing and summarizing our conclusions in Section \ref{sec:conclusion}. We use natural units throughout this work, $\hbar = c= k_B =1$ and the reduced Planck mass $M_P =  2.435 \times 10^{18}$ \si{\giga \electronvolt}.

\section{Thermal inflation}\label{sec:thermal_inflation}

Let us consider a scalar field $\phi$ interacting with a thermal radiation bath  at temperature $T$, with energy density $\rho_R = \frac{\pi^2}{30} g_* T^4$, where $g_*$ denotes the number of relativistic degrees of freedom. For concreteness, we consider a radiation bath made up of $N_F$ Dirac fermion species $\psi_i$, which interact with the scalar field through Yukawa interactions with universal coupling constant $g$:
\begin{equation}\label{eq:fermions_yukawa}
\mathcal{L}_Y = - g \phi \sum_{i=1}^{N_F} \bar{\psi}_i \psi_i~.
\end{equation}
We take the mass of the fermions $m_{\psi_i} \ll T$, so that they can be treated as relativistic degrees of freedom, but such that $m_{\psi_i} > H$, so that flat quantum field theory calculations for the decay width of scalars into fermions are valid \cite{Bastero-Gil:Initial_Conditions_Inflation}.

We assume that the scalar field $\phi$ corresponds to a renormalizable flat direction, or {\it flaton} field, common in several SUSY/supergravity scenarios \cite{Lyth:Gut_Higgs,Lyth:Thermal_Inflation,Barreiro:aspects_Thermal_Inflation, Dine:Flat_Directions_Baryogenesis, Dvali:gauge_hierarchy}, such that its potential is only lifted by soft terms (such as a mass term from SUSY breaking), and non-renormalizable terms. We are interested in the case where the squared mass term is negative, such that the field acquires a large expectation value $M_0$ at zero temperature from the latter's interplay with the non-renormalizable operators. The interaction with the radiation bath induces, however, a thermal mass correction such that the field's effective mass is of the form \cite{Dolan:symmetry_finite_temperature}:
\begin{equation}
m_{\text{eff}}^2 = \alpha^2 T^2 - m^2~,
\end{equation}
where $m$ corresponds to the zero temperature (tachyonic) mass and $\alpha$ is the effective coupling to the thermal bath. For the Yukawa interactions described above we have  $\alpha^2 = g^2 N_F / 6$ at one-loop order. This implies, in particular, that for temperatures above the critical value, $T_c\equiv m/\alpha$, the origin is a stable minimum of the scalar potential, whereas for lower temperatures the minimum is non-trivial and asymptotes to $M_0$ in the limit of vanishing temperature. The origin thus constitutes a false vacuum state, near which we may write the scalar potential as:
\begin{equation}\label{eq:potential}
V(\phi) = \frac{1}{3} M_0^2 m^2 + \frac{1}{2} m_{\text{eff}}^2  \phi^2 + \cdots~,
\end{equation}
where for concreteness we have chosen the constant term such that, if the leading non-renormalizable term is $\sim \phi^6$ the cosmological constant vanishes at the minimum, $V(\phi=M_0)=0$, although this is not crucial to our analysis. For typical flat directions, $M_0\gg m$, since the scale at which the non-renormalizable operators become relevant is generically large (around the grand unification or even the Planck scale).

If, after the first period of slow-roll inflation, the Universe is reheated to attain a temperature $T>T_c$, the flaton field will thus be driven to the false minimum at the origin by Hubble friction, where it is trapped and gives a contribution $V_0= M_0^2m^2/3$ to the vacuum energy. Since the temperature drops as the universe expands, i.e.~$\rho_R\propto a^{-4}$, eventually this vacuum energy may become dominant, thus triggering a new period of inflation, with expansion rate $H\simeq m M_0/3 M_P \lesssim m$. Thermal inflation thus begins when the temperature drops below:
\begin{equation}\label{eq:initial_temperature}
T_i = \bigg(\frac{10}{g_* \pi^2}\bigg)^{\frac{1}{4}} \sqrt{M_0 m}~.
\end{equation} 
Assuming that there is no significant entropy production during thermal inflation, as we confirm in Appendix \ref{app:thermodynamical_considerations}, the temperature of the radiation bath drops as $T \propto a^{-1}$ during thermal inflation, eventually reaching the critical value $T_c$ below which the minimum at the origin is destabilized. The nature of the phase transition (or smooth crossover) that ensues is model-dependent and irrelevant to our discussion (see e.g. \cite{Yamamoto:phase_transition_superstring}), since we are mostly interested in what happens for temperatures $T_c<T<T_i$.

We note that thermal inflation is only possible if the flaton field has a non-negligible interaction with the thermal bath, and in particular $T_i>T_c$ imposes:
\begin{equation}\label{eq:bound_on_alpha}
	\alpha > \bigg( \frac{g_* \pi^2}{10}\bigg)^{\frac{1}{4}} \sqrt{\frac{m}{M_0}}~.
\end{equation} 
For instance, for $m\sim 10$ TeV and $M_0\sim M_P$, this imposes the lower bound $\alpha\gtrsim 10^{-7}$ for $g_*=10-100$. Although this may not seem too restrictive, we note that the number of e-folds of thermal inflation is given by:
\begin{equation}\label{eq:thermal_inflation_duration}
	N_e^{\text{(TI)}} = \ln \bigg( \frac{T_i}{T_c} \bigg) =\frac{1}{2} \ln \bigg(\frac{M_0}{m}\bigg) + \frac{1}{4} \ln \bigg(\frac{10}{\pi^2 g_*}\bigg) + \ln(\alpha)~.
\end{equation}
For the reference values given above, we see that a period of thermal inflation lasting more than 10 e-folds is only possible for $\alpha\gtrsim 0.01$, with even larger effective couplings required for scenarios with a smaller hierarchy between the mass scales $m$ and $M_0$.

We note that inflation does not necessarily end when the temperature falls below $T_c$, since expansion only stops accelerating once the flaton's kinetic energy surpasses its potential energy. Below $T_c$ the field develops a tachyonic instability, since $m_{\text{eff}}^2\simeq -m^2<0$ once $T \ll T_c$, and its value moves away from the origin as $\sim e^{mt}\sim e^{{\frac{m}{H}}N_e}$ for $H\lesssim m$, and there may be a period of {\it fast-roll inflation} \cite{Linde:2001ae} until the field gets close to the minimum at $M_0$ and its kinetic energy takes over. Note that, in the opposite regime $m\lesssim H$, thermal inflation would be followed by an additional period of {\it slow-roll} inflation, but we will not consider this regime in our discussion. The duration of the fast-roll period is, of course, model dependent and, moreover, dependent on the mean field value at the critical temperature.

 In \cite{Dimopoulos:thermal_inflation_primordial_black_holes, Linde:2001ae} it was shown that this period may last for as much as, or even longer than, the thermal inflation period for $H/m\lesssim 1$, depending on the flaton's mass value. This assumed, however, that the mean field value at the critical temperature is set by quantum fluctuations, which as we will see is not necessarily the case. In particular, thermal fluctuations typically enhance the field's variance at $T_c$, therefore reducing the duration of the subsequent fast-roll period. For this reason, we will restrict our analysis to the thermal inflation period ($T_c<T<T_i$), discussing the implications of our results to the subsequent cosmological evolution at the end of our discussion.

Independently of whether or not there is a significant period of inflation below $T_c$, the field will eventually begin oscillating about the minimum of its potential and decay away through the Yukawa interactions in Eq.~(\ref{eq:fermions_yukawa}) \cite{Kofman:reheating_after_inflation}. Although we do not specify the exact nature of the fermion fields in the thermal bath, since we are only modelling the interactions between the flaton and the ambient radiation and our discussion is largely independent of the particular interactions considered, it is implicit that such interactions will eventually lead to the reheating of the Standard Model degrees of freedom at temperatures exceeding at least a few MeV to ensure the correct conditions for primordial nucleosynthesis.

We note that having late thermal inflation and fast-roll inflation periods alters the predictions of inflationary cosmology \cite{Dimopoulos:thermal_infation_minimal_hybrid_inflation}, since the largest CMB scales leave the horizon 50-60 e-folds before the end of the full inflationary epoch, including the primary slow-roll inflation period, which therefore must necessarily be shorter.

Although the leading effect of the interactions between the flaton and the thermal bath is the thermal mass correction responsible for its trapping at the origin, it also induces fluctuation-dissipation effects in the flaton's dynamics that, as we will see, can play an important role in the evolution of field perturbations during thermal inflation\footnote{Note that, in warm inflation, the thermal corrections to the scalar potential need to be suppressed by some mechanism so that dissipative effects play the dominant role in the corresponding slow-roll dynamics. On the other hand, in the case of thermal inflation that we are considering in this work, the dissipative effects are crucial in stabilizing the flaton field at the false minimum.}. These have been considered in \cite{Hiramatsu:Thermal_Fluctuations_Thermal_Inflation} to analyze the nature of the phase transition at $T_c$, but their effects on the associated spectrum of curvature perturbations have so far been overlooked. To study them, we consider the full Langevin-like equation for the flaton field modes $\phi_k$ of comoving momentum $k$ in the spatially flat gauge, which can be obtained through standard techniques in linear response theory assuming the ambient radiation bath is close to an equilibrium state, and is given by (see e.g.~\cite{Berera:Warm_Microphysical, Calzetta:Nonequilibrium_QFT}):
\begin{equation}\label{eq:langevin equation}
\ddot{\phi}_k + (3 H + \Gamma_\phi) \dot{\phi}_k + \omega_k^2 \phi_k = \xi_k~,
\end{equation}
where $\omega_k^2 = k^2/a^2 + m_\text{eff}^2$ and $\Gamma_\phi$ is the dissipation coefficient, which for a field oscillating near a local minimum of its potential (in this case the false minimum at the origin for $T>T_c$) coincides with its finite-temperature decay width \cite{Moss:2008lkw, Yokoyama:2004pf}\footnote{In the spatially flat gauge the temporal component of the metric perturbation does not vanish, but this has no significant effect on the field fluctuations equation to leading order. Since the background field is essentially sitting at the origin during thermal inflation, i.e. $\phi \sim \dot{\phi} \sim \ddot{\phi} \sim 0$ , then the field equation for the flaton perturbations will reduce to the form \eqref{eq:langevin equation}  (c.f. Appendix B, Eq. (B16) in~\cite{Das:2020xmh}).}. We give a brief overview of the derivation of the Langevin-like equation in Appendix \ref{app:Langevin}. On the right hand side of \eqref{eq:langevin equation}, $\xi_k$ is a stochastic noise term which encodes the randomness of the field's interactions with the thermal bath. For modes with physical momentum $p=k/a \lesssim \pi T$ it is well approximated by a gaussian white noise term with a two-point correlator given by the fluctuation-dissipation relation \cite{Hiramatsu:Thermal_Fluctuations_Thermal_Inflation, Gleiser:Microphysical_approach}:
\begin{equation}\label{eq:thermal noise correlator}
\braket{\xi_{k}( t) \xi_{{k}'}( t')} = 2 \Gamma_\phi T  \frac{(2 \pi)^3}{a^{3}} \delta^3(\bm{k} + \bm{k}') \delta (t - t')~.	
\end{equation} 
We note that physically this is reminiscent of the Brownian motion of a heavy particle in an gas, for which random collisions with the gas molecules induce an effective friction that damps its motion. However, the particle never actually comes to rest due to the very same random collisions, eventually reaching an equilibrium with the gas. We expect something very similar to occur to the flaton field modes, with the combined effects of dissipation ($\Gamma_\phi$) and thermal fluctuations ($\xi_k$) driving the field towards a thermal equilibrium with the radiation bath. This behaviour has been observed for scalar fields interacting with a radiation bath both in an inflationary and non-inflationary context \cite{ Bastero-Gil:Initial_Conditions_Inflation, Rosa:2021gbe}, so we anticipate that the same will occur in the case of thermal inflation.

At finite temperature the flaton decay width into relativistic  fermions is given by \cite{Bastero-Gil:2010dgy, Bastero-Gil:Initial_Conditions_Inflation}:
\begin{equation}\label{eq:decay_width_full}
\Gamma_\phi (p) = \frac{3 m_{\text{eff}}^2 \alpha^2}{4 \pi \omega_p} \bigg\{ 1 + \frac{2 T}{p} \ln \bigg[ \frac{1 + \exp(-\frac{\omega_+}{T})}{1 + \exp(-\frac{\omega_-}{T})}\bigg]\bigg\}~,
\end{equation}
where $ \omega_\pm = \frac{|\omega_p \pm p|}{2}$ and we have neglected the mass of the fermions, $m_{\psi_i} \ll T$. Note that fermions acquire a mass through their interaction with the flaton field but, as we will obtain bellow, $ \sqrt{\braket{\phi^2}} \lesssim T$ for perturbative couplings. 

Since the thermal bath will excite field modes $p \lesssim T$ and $m_{\text{eff}} \lesssim T$, the decay width can be well approximated by:
\begin{equation}\label{eq:decay_width_approx}
\Gamma_\phi \simeq \frac{3 m_{\text{eff}}^2 \alpha^2}{16 \pi T}\simeq {\frac{3\alpha^4}{16\pi}} T~,
\end{equation}
where in the last step we have used $m_{\text{eff}}\simeq \alpha T$ for $T\gtrsim T_c$. At the onset of thermal inflation, we then have:
\begin{eqnarray}\label{eq:Gamma_H_initial}
\left.{\frac{\Gamma_\phi}{H}}\right|_{T_i}&\simeq &{\frac{9}{16\pi}}\left({\frac{10}{g_*\pi^2}}\right)^{1/4}\alpha^4 {\frac{M_P}{\sqrt{M_0m}}}\nonumber\\
&\simeq& 2.3 g_*^{-1/4}\left({\frac{\alpha}{0.03}}\right)^4\left({\frac{M_P}{M_0}}\right)^{1/2}\left(\frac{m}{10\ \mathrm{TeV}}\right)^{-1/2}~,
\end{eqnarray}
so that we expect dissipative effects to play an important role in the field's dynamics roughly for the same range of the effective coupling $\alpha$ leading to a period of thermal inflation lasting for more than 10 e-folds, as we have seen above. In the next section we compute the thermal field correlators and associated curvature perturbation power spectrum to better quantify this statement.   

\section{Curvature Perturbations}\label{sec:curvature}

Let us consider the gauge-invariant curvature perturbation on uniform density hypersurfaces, which in the spatially flat gauge can be written as \cite{Bardeen:spontaneous_creation,Baumann:inflation_Lectures}:
\begin{equation}\label{eq:zeta}
	\zeta = - \frac{H}{\dot{\braket{\rho}}} \delta \rho~,
\end{equation}
where the perturbation of a generic function is given by $\delta f(t,\bm{x}) \equiv f(t, \bm{x}) - \braket{f(t, \bm{x})}$, and brackets denote its thermal averaged value. The dimensionless power spectrum of $\zeta$ is defined as  \cite{Dimopoulos:thermal_inflation_primordial_black_holes}:
\begin{equation}\label{eq:power_spectrum}
	\begin{aligned}
		\Delta^2_{\zeta} (k) &= \frac{k^3}{2 \pi^2}  \int d^3 x \; \exp(-	i \bm{k} \cdot \bm{x}) \braket{ \zeta (0) \zeta (\bm{x})}~,\\
		&=  \frac{2k^3}{(2 \pi)^2} \bigg( \frac{H}{ \dot{\braket{\rho}}}\bigg)^2 \int d^3 x \; \exp(-	i \bm{k} \cdot \bm{x}) \braket{ \delta \rho (0) \delta \rho (\bm{x})}~.
	\end{aligned}
\end{equation}
The total energy density $\rho$ during thermal inflation includes the contributions from both the flaton field and the radiation fluid \cite{Kolb:early}:
\begin{equation}\label{eq:energy_density_total}
\rho = \rho_\phi + \rho_R =   \frac{1}{2} \dot{\phi}^2 + V(\phi)+ \frac{1}{2} a^{-2}(t) \partial_i \phi \partial_i \phi+ \frac{\pi^2}{30} g_* T^4~,
\end{equation}
and so we have:
\begin{subequations}
\begin{align}
	\braket{\rho} &= \frac{\pi^2}{30}  g_* T^4 + \frac{1}{3} m^2 M_0^2 + \frac{1}{2} m_{\text{eff}}^2 \braket{\phi^2} + \frac{1}{2} \braket{\dot{\phi}^2} + \frac{1}{2} a^{-2} \braket{\partial_i \phi \partial_i \phi} \label{eq:total_energy_density_averaged}~, \\
	\delta \rho &= \frac{1}{2} m_{\text{eff}}^2 \delta (\phi^2) + \frac{1}{2} \delta (\dot{\phi}^2) + \frac{1}{2} a^{-2} \delta(\partial_i \phi \partial_i \phi) \label{eq:delta_rho}~.
\end{align}
\end{subequations}
Since density perturbations involve perturbations of quadratic functions of the field and its derivatives, the power spectrum, Eq. \eqref{eq:power_spectrum}, involves contributions of the form:
\begin{equation}\label{eq:perturbations_generic_correlation_1}
\braket{\delta(X_i(0)^2) \delta (X_j(\bm{x})^2)} = \braket{X_i(0)^2 X_j(\bm{x})^2} - \braket{X_i(0)^2} \braket{X_j(\bm{x})^2}~,
\end{equation}
where $X_i$ generically denotes the field perturbations and their derivatives. The first term on the right-hand side corresponds to 4th moments involving the gaussian variables $X_i$. According to Isserlis' theorem \cite{Isserlis:Isserlis_Theorem} it is possible to write a $k$th moment of zero-average gaussian variables in terms of their variances. Thus, the correlators can be simply written as \cite{Vignat:Isserlis_theorem}:
\begin{equation}\label{eq:perturbations_generic_correlation_2}
\braket{\delta(X_i(0)^2) \delta (X_j(\bm{x})^2)} = 2 \braket{X_i(0) X_j(\bm{x})}^2~.
\end{equation}
The two-point correlation function for the energy density is then:
\begin{equation}\label{eq:energy_density_perturbations_variance}
\begin{aligned}
	\braket{\delta \rho(0) \delta \rho(\bm{x})} &= \frac{m_\text{eff}^4}{2}  \braket{\phi(0) \phi(\bm{x})}^2 + m_\text{eff}^2 \braket{\phi(0) \dot{\phi}(\bm{x})}^2 + a^{-2} m_\text{eff}^2 \braket{\phi(0) \partial_i\phi(\bm{x})}^2~,\\
	& \quad + \frac{1}{2} \braket{\dot{\phi}(0) \dot{\phi}(\bm{x})}^2 + a^{-2} \braket{\dot{\phi}(0) \partial_i\phi(\bm{x})}^2 + \frac{a^{-4}}{2}  \braket{\partial_i\phi(0) \partial_j\phi(\bm{x})}^2~,
\end{aligned}
\end{equation}
that is, contributions from all possible correlation functions involving $\phi$, $\dot{\phi}$ and $\partial_i \phi$.

We note that we are interested in computing the curvature perturbation power spectrum on super-horizon scales, $k\ll aH$. To do this we need to compute the field variance $\langle \phi^2\rangle$ and the average kinetic and gradient energies appearing in Eq.~(\ref{eq:total_energy_density_averaged}), which involve integrating over all thermally excited field modes\footnote{Note that the field variance and related correlation functions exhibit the usual zero-temperature divergences that need to be appropriately renormalized. The finite temperature contributions that we are interested in are, however, finite and thus require no additional renormalization procedure (see e.g.~\cite{Bellac:thermal_field}).}. However, the correlation function of the noise terms is exponentially suppressed for length scales smaller than the correlation length $\xi \sim (\pi T)^{-1}$, that is for physical momentum scales larger than $p\sim \pi T$ \cite{Hiramatsu:Thermal_Fluctuations_Thermal_Inflation}. Therefore, we may consider the hard cutoff, $p = \pi T$, as a good approximation when computing \eqref{eq:energy_density_perturbations_variance}. This can be translated into a comoving momentum cutoff $k_c=\pi T_c$ if we set $a(T_c)=1$, following the conventions of \cite{Dimopoulos:thermal_inflation_primordial_black_holes} to allow for a better comparison with the purely quantum calculation.   

To compute the power spectrum we need the three combinations of the correlations between $\phi_k$ and $\dot{\phi}_k$, i.e. $\braket{\phi_k \phi_k}$, $\braket{\phi_k \dot{\phi}_k}$ and $\braket{\dot{\phi}_k \dot{\phi}_k}$. These are the building blocks of all the remaining correlation functions involved in the power spectrum. We will explicitly compute the correlator of the field modes and list all others in Appendix \ref{appendix:correlations} as their computation follows similar steps. 

The equal-time two-point correlation function of the field modes can be written in terms of the Green's function associated with \eqref{eq:langevin equation} and the noise correlator:
\begin{equation}\label{eq:field_mode_variance}
\braket{\phi_k (z) \phi_{k'} (z)} = H^{-4} \int_{z_i}^{z} ds_1 \; \int_{z_i}^{z} ds_2 \; s_1^{-2} s_2^{-2}  G_s (z, s_1)  G_s(z, s_2) \braket{\xi_{k} (s_1) \xi_{k'} (s_2)}~,
\end{equation}
where we have traded the time-dependence for a dependence on the variable $z = T/H$, with $z_i = T_i/H$. Note that $z\propto a^{-1}$ during thermal inflation, so that it is a decreasing function of time. For simplicity, we have chosen initial conditions such that the thermal field correlator vanishes at the onset of thermal inflation (the full correlator will, of course, include quantum fluctuations as we discuss in Section 4), which constitutes a “worst-case scenario” as we consider below.  Note that formally the homogeneous (noise-independent) solutions of \eqref{eq:langevin equation} should be included, and that these depend on whether the field is displaced from the high-temperature minimum at the origin after the first slow-roll inflation period. Nevertheless, such a displacement is typically damped away well before thermal inflation begins, so that we may safely discard the contribution from the homogeneous solutions. These solutions are required, however, to compute the Green's function, which is given by the usual expression:
\begin{equation}\label{eq:green_function_definition}
G_s (z, s) = \frac{ \phi_k^{(1)}(s) \phi_k^{(2)}(z) -  \phi_k^{(1)}(z)  \phi_k^{(2)}(s)}{W( \phi_k^{(1)}, \phi_k^{(2)})(s)}~,\\
\end{equation}
where $\phi_k^{(1)}$ and $\phi_k^{(2)}$ are the homogeneous solutions of equation \eqref{eq:langevin equation} and $W$ denotes their Wronskian.

During most of thermal inflation, except for temperatures close to the critical value, the thermal mass dominates over the field's zero temperature mass, $\alpha T \gg m$. This allows us to compute analytically the field modes, and thus obtain the field’s two-point correlation function with a decay width of the form \eqref{eq:decay_width_approx}.

The homogeneous equation of motion for the flaton field modes \eqref{eq:langevin equation} can be written in terms of the $z$ variable as:
\begin{equation}\label{eq:langevin_equation_z_variable_thermal_mass_domination}
z^2 \phi_ k'' -z \left( 2 + \gamma z\right) \phi_k' + z^2\bar\omega_k^2  \phi_k = 0~,
\end{equation}
where $\bar\omega_k^2\equiv \omega_k^2 /T^2 \simeq k^2/T_c^2 + \alpha^2$ and $\gamma\equiv 3\alpha^4/16\pi$, such that $\Gamma_\phi/H=\gamma z$. Let us define $\phi_k= z e^{\gamma z/2}\chi_k$, such that:
\begin{equation}\label{eq:langevin_equation_z_chi}
\chi_ k'' +\left(\bar\omega_k^2 -{\gamma^2\over4}-{\gamma\over z}-{2\over z^2}\right) \chi_k = 0~.
\end{equation}
Even though we can express the exact solutions of the above equation in terms of Whittaker functions \cite{Abramowitz:Handbook}, it is more instructive to note that, since $\gamma\ll \bar\omega_k^2$ for $\alpha\lesssim 1$ and $z>z_c=m/\alpha H >\alpha^{-1}>1$, we may neglect all the terms inside the brackets in Eq.~(\ref{eq:langevin_equation_z_chi}) except for the one involving $\bar\omega_k^2$ to a good approximation. This means that the homogeneous solutions are approximately given by:
\begin{equation}\label{hom_sol}
\phi_k^{(1)}(z)\simeq ze^{{\gamma\over2} z}\sin(\bar\omega_k z)~, \qquad \phi_k^{(2)}(z)\simeq ze^{{\gamma\over2}  z}\cos(\bar\omega_k z)~,
\end{equation}
thus constituting oscillatory functions in the $z$ variable with an amplitude decreasing due to both Hubble expansion ($z\propto a^{-1}$) and the field's decay into the light fermions. This yields the Green's function:
\begin{equation}\label{eq:green_function_thermal_mass_domination}
G_s (z, s) ={1\over \bar\omega_k} \frac{z}{s}  \exp \bigg[{\gamma\over 2} (z-s)\bigg] \sin \bigg[  \bar\omega_k (z-s) \bigg]~.
\end{equation}
The noise correlation function \eqref{eq:thermal noise correlator} can be written in terms of the $z$ variable as:
\begin{equation}\label{eq:noise_variance_in_z}
\begin{aligned}
	\braket{\xi_{k}(z_1) \xi_{k'}(z_2)} &= 2 H z_1 \Gamma_\phi T  \frac{(2 \pi)^3}{a^{3}} \delta^3(\bm{k} + \bm{k}') \delta (z_1 - z_2)~,\\
	&\simeq 2\gamma H^3z_1^6 \frac{(2 \pi)^3}{z_c^3} \delta^3 (\bm{k} + \bm{k}') \delta( z_1 -z_2)~,\\
\end{aligned}
\end{equation}
where in the second line we used the dominance of the thermal mass for $T>T_c$.

We may now substitute Eqs.~\eqref{eq:green_function_thermal_mass_domination} and \eqref{eq:noise_variance_in_z} into Eq.~\eqref{eq:field_mode_variance} to obtain the field's two-point correlation function:
\begin{equation}\label{eq:field_mode_correlator}
\braket{\phi_k (z) \phi_{k'} (z)} = (2 \pi)^3 \delta^3(\bm{k} + \bm{k}') \frac{T}{a^3 \omega_k^2}(1-\delta)~, \qquad \delta=  \exp \bigg[ - \frac{3 \alpha^4}{16 \pi} \frac{T_i}{H} \bigg(1 - \frac{T}{T_i} \bigg)\bigg]~,
\end{equation}
where again we used that $\bar\omega_k\gg \gamma$. Note that for $\Gamma_\phi/H (T_i)\gtrsim 1$, we have $\delta\ll 1$ for all temperatures below $T_i$ (but above $T_c)$, thus yielding a thermal equilibrium distribution for the field modes that is independent of the decay width. This means that if the field decays efficiently at the onset of thermal inflation it will attain an equilibrium distribution that simplify redshifts with expansion (with corresponding decrease in temperature). This is a generic result\footnote{This was already known in the simple case of a one-dimensional harmonic oscillator in Brownian motion \cite{Chandrasekhar:review}. Here also the late time variance of the displacement ceases to depend on the viscosity of the fluid, as the suspended particle has reached equilibrium with its environment.} obtained in other cosmological contexts \cite{ Bastero-Gil:Initial_Conditions_Inflation, Rosa:2021gbe} that we now recover also within thermal inflation -- it simply states that if the field interacts significantly with the thermal bath at some point during its evolution it reaches a near-thermal configuration that is subsequently maintained unless there is some significant change in the field's properties (in our case the tachyonic instability just below the critical temperature).

We note that the two-point correlation function vanishes at the onset of thermal inflation by construction, since the integral Eq.~(\ref{eq:field_mode_variance}) is zero at $z=z_i$. This assumes that field modes were not excited when thermal inflation begins, which need not be the case since interactions with the thermal bath are present in the prior radiation-dominated epoch. If field modes thermalize before its vacuum energy becomes dominant, Eq.~(\ref{eq:field_mode_correlator}) will nevertheless hold (with $\delta\simeq 0$), since this result is also valid for a radiation-dominated cosmological background \cite{Bastero-Gil:Initial_Conditions_Inflation}. However, we note that during the radiation era $\Gamma_\phi/H \propto T/H \propto a$, while $\Gamma_\phi/H\propto a^{-1}$ during thermal inflation, so that this ratio attains its maximum value at the onset of thermal inflation. Recalling Eq.~(\ref{eq:Gamma_H_initial}), we conclude that $\alpha\gtrsim 0.01$ is required for field thermalization if the zero temperature mass $m$ is not far from the TeV scale at which new physics may be expected. As discussed in the previous section, this is exactly the regime where a period of thermal inflation lasting more than 10 e-folds (and which can in particular sufficiently dilute unwanted relics of the first reheating process) can occur. We will thus henceforth focus our analysis on this parametric regime, in which the field thermalizes either before or at the onset of the thermal inflation epoch.

We may now use Eq.~\eqref{eq:field_mode_correlator} to compute the field variance and related correlation functions, as we detail in Appendix \ref{appendix:correlations}. We obtain for the total average energy density:
\begin{equation}\label{eq:averaged_rho}
\braket{\rho} = \frac{\pi^2}{30}  \left(g_* + \frac{5}{\pi} (1-\delta)\right)T^4 + \frac{1}{3} m^2 M_0^2~,
\end{equation}
where we note that the field contributes essentially as an additional bosonic degree of freedom to the radiation energy density if thermalization is efficient ($\delta \ll 1$). Its contribution is not exactly one degree of freedom since we have considered a hard cutoff on the momentum of the modes that are excited by interactions with the thermal bath at $k_c=\pi T_c$. This is only an approximation to the smooth cutoff associated with the noise correlator \cite{Hiramatsu:Thermal_Fluctuations_Thermal_Inflation}, which nevertheless captures the essential physics of the problem.

Using the values of each component of the power spectrum \eqref{eq:perturbations_generic_correlation_1} given in Appendix \ref{appendix:correlations}, the density perturbations are:
\begin{equation}\label{eq:density_perturbations}
\int d^3 x \; \exp(-	i \bm{k} \cdot \bm{x}) \braket{\delta \rho (0) \delta \rho (\bm{x})} \approx \frac{\pi T^5}{6 a^3} \bigg[1 + 3 \bigg(\frac{3 \alpha^4}{32 \pi^2}\bigg)^2 \bigg(1 - \frac{\alpha}{\pi} \arctan\left({\pi \over\alpha}\right)\bigg) \bigg] (1-\delta)^2~,
\end{equation}
to leading order on super-horizon scales $k< aH < \alpha T_c$. We note that the first term within the square brackets dominates over the second one. This then yields for the power spectrum on super-horizon scales:
\begin{eqnarray}\label{eq:thermal_power_spectrum}
{\Delta^2_{\zeta}}^{\text{(therm)}} (k) &=& \frac{150}{(2 \pi)^5} \frac{k^3}{T_c^3} \frac{(1-\delta)^2}{\bigg[g_* + \frac{5}{\pi} (1-\delta) -\frac{5}{\pi} \frac{3 \alpha^4}{64 \pi} \frac{T}{H} \delta \bigg]^2}\nonumber~,\\
&\simeq&  \frac{150}{(2 \pi)^5} \frac{\alpha^3}{g_{*, \phi}^2} \bigg(\frac{H}{m}\bigg)^3 \bigg(\frac{k}{k_c}\bigg)^3~,
\end{eqnarray}
where in the second line we have taken the prompt thermalization limit, i.e.~$\delta\ll 1$, in which case the flaton field contributes to the total number of relativistic degrees of freedom, given by $g_{*, \phi} \simeq  g_* + 5/\pi$. Note that this result is time-independent, reflecting the freeze-out of curvature perturbations on super-horizon scales and thus the single-fluid nature of the dynamics, i.e.~the fact that the flaton field thermalized with the radiation bath. 

The power spectrum is blue-tilted so its maximum value is attained for the last scale to leave the horizon during thermal inflation, i.e. $k_c = H$ which leaves at $T=T_c$. Although our calculation assumes the dominance of the thermal part of the flaton's mass, an approximation that breaks down close to the critical temperature, we may extrapolate our results with a reasonable accuracy to $k_c$, thus yielding an upper bound on the power spectrum of scales leaving the horizon before the phase transition, in the thermal equilibrium limit:
\begin{eqnarray}\label{eq:thermal_power_spectrum_max}
{\Delta^2_{\zeta}}^{(\text{therm, max)}} (k)\simeq  \frac{150}{(2 \pi)^5} \frac{\alpha^3}{g_{*, \phi}^2} \bigg(\frac{H}{m}\bigg)^3~.
\end{eqnarray}
The power spectrum would, thus, be maximized for $g_{*, \phi} \sim \alpha \sim \frac{H}{m} \sim 1$, yielding ${\Delta^2_{\zeta}}^{(\text{therm, max)}} \sim10^{-2}$, but in realistic scenarios with perturbative couplings and at least one fermionic degree of freedom in the ambient thermal bath the power spectrum should have a parametrically smaller amplitude. 

Hence, if the flaton field has significant interactions with the radiation bath, $\alpha\gtrsim 0.01$ (as expected in scenarios with a significant number of e-folds of thermal inflation), the thermal nature of its fluctuations suppresses the amplitude of the induced curvature perturbations on super-horizon scales, which is the main result of this work. While this may seem surprising, given that thermal fluctuations generically have a larger amplitude than quantum vacuum fluctuations (as considered in \cite{Dimopoulos:thermal_inflation_primordial_black_holes}), it has a simple physical explanation: fluctuation-dissipation effects increase not only the density fluctuations on super-horizon scales but also the field variance and the average gradient and kinetic energies, thus, the average energy density. The latter effect turns out to be more significant and, hence, decreases the amplitude of the associated curvature power spectrum with respect to the quantum case.

A relevant consequence of our analysis is that, in realistic scenarios, we do not expect the amplitude of the curvature power spectrum to be sufficiently large to lead to the formation of primordial black holes, which would require $\Delta^2_\zeta\gtrsim 10^{-2}$ \cite{Green:constraints_pbh,Khlopov:pbh,Green:sirens_pbh}, at least on scales that become super-horizon above the critical temperature. This motivates a better comparison with the results obtained in \cite{Dimopoulos:thermal_inflation_primordial_black_holes} for quantum flaton fluctuations, where larger curvature perturbations were obtained. We pursue this comparison in the next Section.

\section{Comparison between the thermal and quantum power spectra}\label{sec:comparison}

The linear approximation to the quantum power spectrum is given in \cite{Dimopoulos:thermal_inflation_primordial_black_holes} by\footnote{This was computed considering only the contribution from super-horizon field modes, effectively removing the unphysical divergence associated with sub-horizon modes. Note that the thermal contribution is finite, i.e., that the only divergent piece arises in the zero-temperature quantum contribution.}:
\begin{equation}\label{eq:quantum_power_spectrum_k_over_kc}
	{\Delta^2_{\zeta}}^{\text{(quan)}} (k)  = \frac{4}{\sqrt{\pi}} \frac{ \Gamma(\nu)}{\nu^2 \Gamma\big(\nu - \frac{3}{2}\big)} \bigg(\frac{H}{m}\bigg)^{3 - 2 \nu} \bigg(\frac{k}{k_c}\bigg)^3 \bigg[\bigg(\frac{k}{k_c}\bigg)^2 + \frac{m^2}{H^2}\bigg]^{-\nu}~,
\end{equation}
where $\nu=\sqrt{m^2/H^2+9/4}$. To better compare our results with those obtained assuming purely quantum flaton fluctuations in \cite{Dimopoulos:thermal_inflation_primordial_black_holes}, we plot both power spectra as a function of comoving momentum in Figure 1. We show the case of $H/m=0.3$ (which according to the analysis in \cite{Dimopoulos:thermal_inflation_primordial_black_holes} yields all dark matter in the form of primordial black holes) and taking  $\alpha=1$, $N_F=1$ and $\delta = 0$ to maximize the thermal power spectrum. We note that the thermal power spectrum is only shown up to $k=k_c$, since our calculation is only valid for modes that exit the horizon before the phase transition; whereas the quantum calculation can be extended to larger momentum, assuming a subsequent period of fast-roll inflation as mentioned earlier. 

\begin{figure}[htbp]
	\centering
	\includegraphics{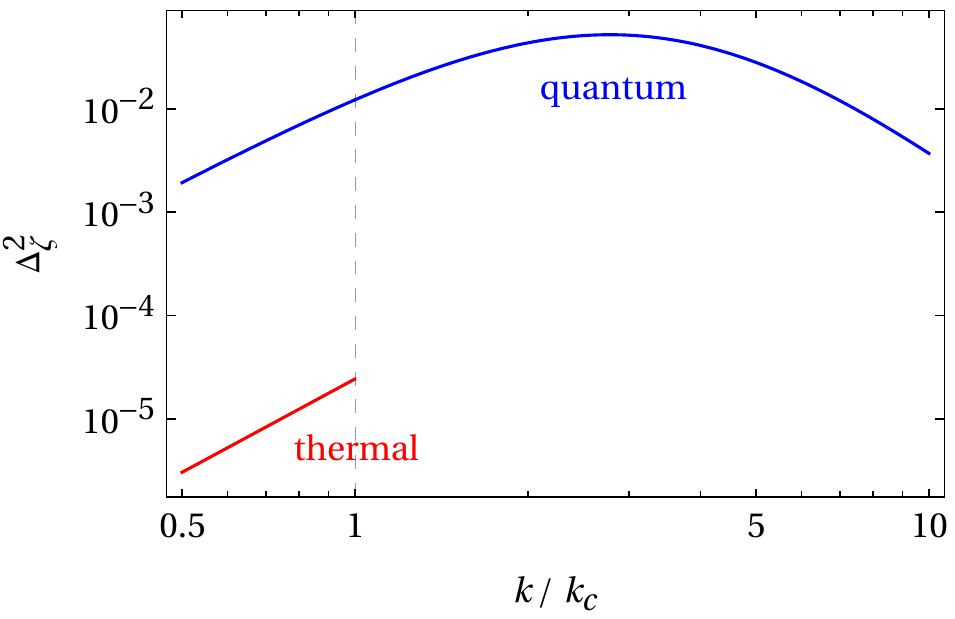}
	\caption{The quantum power spectrum (blue) and the thermal power spectrum (red) as a function of $k$ for $H/m = 0.3$, $\alpha=1$, $N_F=1$ and $\delta = 0$.}
    \label{fig:thermal_quantum_power_spectra_comparison}
\end{figure}

As one can clearly see in this figure, thermal fluctuations significantly suppress the curvature perturbation spectrum with respect to the quantum case, for the reasons explained in the above section. Furthermore, whereas quantum vacuum fluctuations may yield a sufficiently large amplitude that leads to primordial black hole formation, a thermalized flaton field induces much smaller perturbations, although they may nevertheless exceed the even smaller fluctuations observed on large scales in the CMB anisotropies spectrum. 

We should note that the quantum power spectrum peaks at scales that leave the horizon for $T<T_c$, where our approximations break down. Extending our calculation to this regime would involve a different form of the dissipation coefficient, since as the field experiences a tachyonic instability the latter no longer corresponds to the perturbative decay width at finite temperature. Let us note, however, that fluctuation-dissipation effects are more pronounced at the start of thermal inflation as discussed earlier, so that they no longer play a significant role near $T_c$. If the field thermalizes at the onset of thermal inflation, it will nevertheless maintain an equilibrium distribution with a decreasing temperature due to inflationary expansion. Let us then compare the magnitude of field fluctuations at $T_c$ in both the quantum vacuum and thermal cases. The thermal variance is obtained by expanding the field in terms of its modes:
\begin{equation}
	\braket{\phi(\bm{x}) \phi(\bm{y})} = \int \frac{d^3 k}{(2 \pi)^3} \frac{d^3 k'}{(2 \pi)^3} \; \braket{\phi_k \phi_{k'}} \exp(i \bm{k} \cdot \bm{x}) \exp(i \bm{k} \cdot \bm{y})~,
\end{equation}
and using the field modes correlator \eqref{eq:field_mode_correlator}, we obtain for the field variance in the thermalized limit:
\begin{equation}\label{eq:thermal_variance}
	\braket{\phi^2}_{\text{therm}} = \frac{2}{(2 \pi)^2} \frac{T}{a} \int_{0}^{k_\text{cutoff}} dk \; \frac{k^2}{k^2 + \alpha^2 T_c^2} = \frac{ T^2}{2 \pi} \bigg[1 - \frac{\alpha}{\pi}\arctan \bigg(\frac{\pi}{\alpha}\bigg)\bigg]~,
\end{equation}
which we note is only mildly dependent on the effective coupling $\alpha$, while the quantum one is given by \cite{Dimopoulos:thermal_inflation_primordial_black_holes}:
\begin{equation}\label{eq:quantum_variance}
	\braket{\phi^2}_{\text{quan}} = \bigg(\frac{H}{2 \pi}\bigg)^2 \frac{\Gamma^2(\nu) 2^{2 \nu}}{6 \pi} \bigg(\frac{a H}{m}\bigg)^{2 \nu}  F \bigg(\nu, \frac{3}{2}; \frac{5}{2}; - \bigg(\frac{a H}{m} \bigg)^2\bigg)~,
\end{equation}
where $F(a,b,c,z)$ denotes the Hypergeometric function. The field variance in both cases is shown in Figure 2, where we extrapolate the thermal variance beyond the phase transition purely for comparison purposes.

\begin{figure}[htbp]
	\centering
	\includegraphics[scale=1]{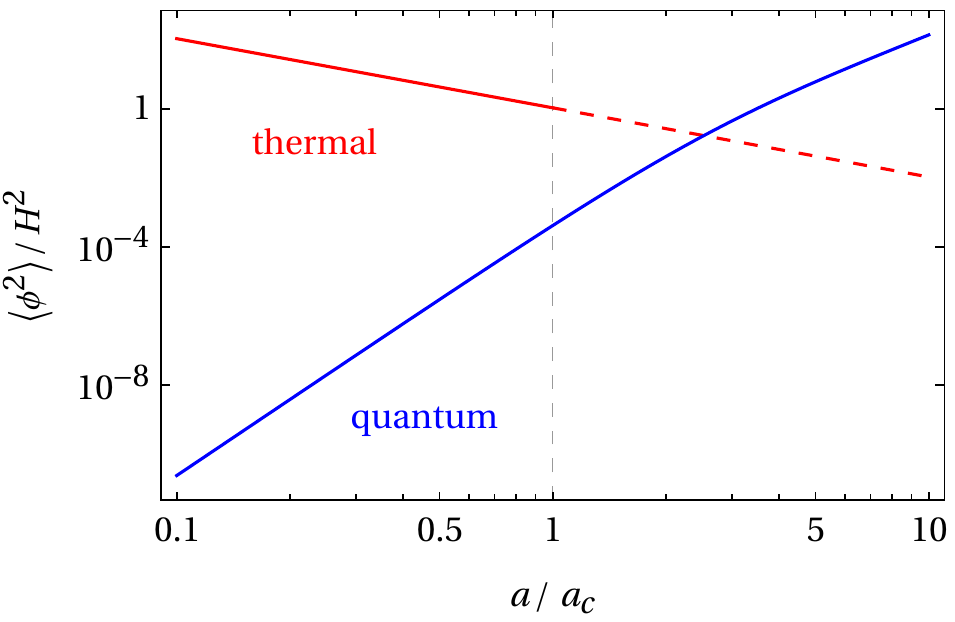}
	\caption{Quantum (blue) and thermal (red) field variance as a function of the scale factor for $H/m= 0.3$, $\alpha =1$ and $\delta = 0$. The critical temperature corresponds to the dashed vertical line, below which the thermal variance is extrapolated, as indicated by the dashed red line.}
	\label{fig:variances_comparison}
\end{figure}
As one can clearly observe in this figure, the quantum field variance is several orders of magnitude smaller than the thermal variance before the phase transition, which validates our calculation in neglecting vacuum fluctuations in the thermalized flaton scenario. While at the critical temperature this is still true, if one extrapolates the thermal variance for $T<T_c$ ($a>a_c=1$), we see that quantum fluctuations become dominant less than one e-fold after the critical temperature is attained. 

While this extrapolation is non-trivial, since the fluctuation-dissipation effects would have to be re-computed, it may suggest that vacuum perturbations may become dominant after the phase transition, in which case the computation in  \cite{Dimopoulos:thermal_inflation_primordial_black_holes} would hold. In fact, the peak in the quantum power spectrum is obtained for modes with $k={H\over 2}\sqrt{3(2\nu+3)}>k_c=H$, which leave the horizon for temperatures below the critical value and thus, in the example shown above, already in the regime where the quantum variance is dominant. 

This would, in fact, suggest that large enough curvature perturbations leading to primordial black hole formation may be generated after thermal inflation (from quantum fluctuations), but it is not clear that quantum and thermal fluctuations may be examined independently nor that the thermal variance maintains its form below $T_c$. In addition, and perhaps most importantly, the fact that the thermal variance is still typically a few orders of magnitude larger than the quantum one at the critical temperature indicates that the flaton field should reach the minimum of its potential much quicker if it thermalizes, therefore considerably shortening, or even possibly, precluding an ensuing period of fast-roll inflation. 

A more complete analysis of the problem including both thermal and quantum fluctuations in the analysis, potentially along the lines of \cite{Ramos:quantum_thermal}, is required to compute the spectrum of curvature perturbations on scales that leave the horizon at temperatures below $T_c$, and is left for future work.

\section{Conclusion}\label{sec:conclusion}

We have computed the spectrum of curvature perturbations generated during thermal inflation taking into account the thermal fluctuations of the flaton field driving this period. These are associated with fluctuation-dissipation effects driven by the flaton's interactions with the ambient radiation bath. Our analysis involved solving the Langevin-like equation effectively describing the evolution of the flaton's Fourier modes. We computed the associated correlation functions in the approximation of a gaussian white noise and a dominant thermal contribution to the flaton's mass, for temperatures above the critical value at which the flaton is held at the false vacuum at the origin. 

We have concluded that, if the flaton's (finite-temperature) decay width exceeds the Hubble parameter at the onset of thermal inflation, the field essentially thermalizes with the ambient radiation bath, contributing approximately as an extra relativistic degree of freedom. This occurs when the effective coupling between the flaton and the thermalized degrees of freedom $\alpha\gtrsim 0.01$, which roughly corresponds to the parametric regime where over 10 e-folds of thermal inflation (above $T_c$) occur. We found that the consequent increase in the field variance and the average gradient and kinetic energies enhances the background energy density (namely its time-dependent part that determines curvature perturbations) with respect to a field with purely quantum vacuum fluctuations analyzed in  \cite{Dimopoulos:thermal_inflation_primordial_black_holes}. Despite the enhancement of super-horizon density fluctuations in the thermal case, the overall amplitude of the curvature power spectrum is significantly reduced with respect to the quantum case, so that thermal fluctuations behave very differently compared to their quantum counterparts, regarding the generation of curvature perturbations during periods of thermal inflation.

While our analysis is not applicable for modes that leave the horizon once the temperature has fallen below the critical value and the field starts rolling towards the true minimum of its potential, we expect thermal effects to become less relevant in this regime and quantum fluctuations to become dominant, potentially yielding large curvature perturbations at such scales as computed in \cite{Dimopoulos:thermal_inflation_primordial_black_holes}. However, a full analysis including both quantum and thermal fluctuations in the dynamics of the flaton field is required to accurately describe the putative fast-roll inflation phase below the critical temperature. It must be noted, in any case, that such a phase is necessarily shortened by the fact that the field variance at the critical temperature, which sets the typical field value at this stage, is much larger if the field thermalizes with the radiation bath. It is therefore unclear whether super-horizon fluctuations with $k>k_c$ can be generated in this phase.

We have modelled the thermal bath through a set of fermion species coupled to the flaton field, but we expect our main conclusions to hold with the inclusion of other bosonic fields, like scalars or vector bosons: if $\Gamma_\phi>H$ at some stage during thermal inflation, the field will be driven towards a thermal fluctuation spectrum. Only the details of how and when this equilibrium is attained may depend on the types of light fields that interact with the flat direction. Our analysis shows that thermalization does not require large coupling constants describing the interaction between the flaton and the radiation bath. In any case such couplings cannot be too suppressed to sustain a sufficiently long period of thermal inflation that may, in particular, dilute any unwanted relics generated after the primary slow-roll inflation period. Hence, fluctuation-dissipation effects cannot in general be neglected in the dynamics of the flaton field and on the curvature perturbations they induce during thermal inflation. This is particularly relevant if one wishes to understand whether thermal inflation periods may leave behind a sizeable population of primordial black holes, and we hope that our work motivates further exploration of these and related issues, including other potential implications for structure formation in our Universe \cite{Leo:structure_formation_thermal_inflation}.

\section*{Acknowledgements}

M.B.G. work has been partially supported by MICINN (PID2019-105943GB-I00/AEI/10.130\\39/501100011033) and “Junta de Andalucía” grant P18-FR-4314. JMG acknowledges the support from the Fundação para a Ciência e a Tecnologia, I.P. (FCT) through the Research Fellowship No. 2021.05180.BD derived from Portuguese national funds. This work was supported by FCT - Fundação para a Ciência e a Tecnologia, I.P. through the projects CERN/FIS-PAR/0027/2021, UIDB/04564/2020 and UIDP/04564/2020, with DOI identifiers 10.54499/CERN/FIS-PAR/0027/2021, 10.54499/UIDB/04564/2020 and 10.54499/UIDP/04\\564/2020, respectively.

\appendix

\section{Derivation of the Langevin-like equation}\label{app:Langevin}

Here we give a brief overview of how to derive the Langevin-like equation (c.f. Eq.~\eqref{eq:langevin equation}) determining the dynamics of a scalar field interacting with an ambient heat bath, which is the basis of our analysis of the dynamics of the flaton field during thermal inflation. We will follow mostly the discussion in \cite{Berera:Warm_Microphysical} and consider, for simplicity, only the flat space case, referring the reader to the detailed discussion in \cite{Berera:Warm_Microphysical} on the inclusion of curved space-time effects, particularly for a flat FLRW universe.

We use the Schwinger-Keldysh formalism, or closed-time path (CTP) formalism of non-equilibrium quantum field theory, to compute the effective action describing the interacting quantum field at finite temperature. In this formalism, the expectation values of operators at a given time $t$ are evaluated using a closed time path, from $t=-\infty$ to $t=+\infty$ and back, and introducing two copies of the same scalar field, $\phi_1$ and $\phi_2$, each with support on each of the branches of the CTP. Field propagators are then described by $2\times2$ matrices $G_{ab}$, where $a,b=1,2$, and similarly for the self-energy matrix $\Sigma_{ab}$. This matrix is, in particular, specified by two functions:
\begin{equation}
i\Sigma_\rho=i(\Sigma_{21}-\Sigma_{12})~,\qquad i\Sigma_F={1\over2}(\Sigma_{21}+\Sigma_{12})~, 
\end{equation}
and it can be shown that in thermal equilibrium they satisfy the Kubo-Martin-Schwinger (KMS) relation (in Fourier space):
\begin{equation}\label{KMS}
 \Sigma_F(\mathbf{p},\omega)=-i\left[n(\omega)+{1\over 2}\right]\Sigma_\rho(\mathbf{p},\omega)~,
\end{equation}
where $n(\omega)$ is the Bose-Einstein distribution (for a bosonic field). The CTP action is of the form:
\begin{equation}
 S[\phi_1,\phi_2]=\int d^4x\left[-{1\over 2}\phi_1\partial^2\phi_1 -V(\phi_1)\right]-(\phi_1\leftrightarrow \phi_2)~.
\end{equation}
It is convenient to use the Keldysh representation $\phi_c=(\phi_1+\phi_2)/2$ and $\phi_\Delta=\phi_1-\phi_2$, such that varying the action with respect to $\phi_\Delta$ and then setting $\phi_\Delta=0$ (since $\phi_1=\phi_2$ correspond to the same field) yields the classical equation of motion $\partial^2\phi_c+V'(\phi_c)=0$.

To obtain the effective equation of motion including the effects of the interactions between the scalar field and fields in the heat bath (specified by some interaction Lagrangian), we first compute the effective action, which incorporates only the effects of 1PI diagrams:
\begin{equation}
e^{i\Gamma[\phi_1,\phi_2]}=\int_{1PI}\mathcal{D}[\phi_1']\mathcal{D}[\phi_2']\rho[\phi_1'(-\infty),\phi_2'(-\infty)]e^{i(S[\phi_1+\phi_1']-S[\phi_2+\phi_2'])}~, 
\end{equation}
where $\rho[\phi_1'(-\infty),\phi_2'(-\infty)]$ denotes the density matrix. The effective action can be expanded in a perturbative series, yielding to quadratic order in $\phi_\Delta$ in the Keldysh representation:
\begin{eqnarray}
\Gamma[\phi_c,\phi_\Delta]&=&-\int d^4x\left[\partial^2\phi_c(x) + V'(\phi_c(x))\right]\phi_\Delta(x)-\int d^4x\int d^4x' \phi_\Delta(x) \Sigma_R(x,x')\phi_c(x')\nonumber\\
&+&{1\over 2}\int d^4x \int d^4x'\phi_\Delta(x)i\Sigma_F(x,x'))\phi_\Delta(x')+ \mathcal{O}(\phi_\Delta^3)~,
\end{eqnarray}
where $\Sigma_R(x,x')=\Sigma_\rho(x,x')\theta(t-t')$ is the retarded self-energy of the field,  including the interactions between the latter and the fields in the heat bath in an explicitly causal manner. By means of a Hubbard-Stratonovich transformation in the path integral, we may replace the quadratic term in the quantum field fluctuations $\phi_\Delta$ by statistical fluctuations of a random field $\xi(x)$:
\begin{eqnarray}
e^{-{1\over2}\int d^4x d^4x'\phi_\Delta(x)\Sigma_F(x,x')\phi_\Delta(x')}=|\det\Sigma_F|^{1/2}\!\!\int \mathcal{D}\xi e^{-{1\over2}\int d^4x d^4x'\xi(x)\Sigma_F^{-1}(x,x')\xi(x')+i\int d^4x \xi(x)\phi_\Delta(x)}~.\nonumber\\
\end{eqnarray}
Thus, varying the effective action with respect to $\phi_\Delta$ and then setting $\phi_\Delta=0$ yields the effective equation of motion for the scalar field:
\begin{eqnarray} \label{Langevin_non_local}
\partial^2\phi_c+ V'(\phi_c)+\int d^4x'\Sigma_R(x,x')\phi_c(x')=\xi(x)~.
\end{eqnarray}
The way the field $\xi(x)$ appears in the path integral shows that it can be interpreted as a Gaussian stochastic noise with zero mean $\langle \xi(x) \rangle=0$ and a two-point correlation function given by the scalar field's self-energy:
\begin{eqnarray}
\langle \xi(x)\xi(x') \rangle = \Sigma_F(x,x')~. 
\end{eqnarray}
Note that, in this formalism, statistical averages over the ensemble characterizing the heat bath are given by functional integration over the $\xi(x)$ field. Due to its zero average, the noise term will not affect the dynamics of the background scalar field, but its non-zero variance will nevertheless source thermal field fluctuations as we analyzed in this work.

The equation of motion (\ref{Langevin_non_local}) includes a non-local term that encodes the dissipative effects of the interactions between the field and the thermal bath, alongside the corrections to the field's mass (both the zero and finite temperature contributions). To see this more explicitly, let us focus on the case of a scalar field oscillating about a minimum of its potential (in the case of interest to our discussion, about the origin when $T > T_c$). For simplicity let us consider only the homogeneous field component (i.e. the super-horizon field modes), of the form $\phi_c=\Phi e^{i\omega t}$, although the discussion can be easily extended to generic Fourier modes (see e.g. \cite{Yokoyama:2004pf}). In this case the non-local term can be written as:
\begin{eqnarray}
\int d^4x'\Sigma_R(x-x')\phi_c(x')&=&\Phi\int_{-\infty}^tdt' \int d^3x'\Sigma_R(x-x')e^{i\omega t'},\nonumber\\    
 &=& \Phi\int_{-\infty}^tdt'\int {d\omega'\over 2\pi} \tilde{\Sigma}_R(\omega',\mathbf{x})e^{i\omega'(t-t')}e^{i\omega t'}~,
\end{eqnarray}
where the spatial integration over $\mathbf{x}'$ is implicit in the definition of $\tilde{\Sigma}_R(\omega',\mathbf{x})$. The time integration can then be performed using  \cite{Yokoyama:2004pf}
\begin{equation}
\int_0^\infty d\tau e^{i(\omega'-\omega)\tau}=iP{1\over \omega'-\omega}+\pi\delta(\omega'-\omega)~,  
\end{equation}
where $P$ denotes the principal value, so that the non-local term becomes:
\begin{eqnarray}
\int d^4x'\Sigma_R(x-x')\phi_c(x')=\Phi \int {d\omega'\over 2\pi} P {i \tilde{\Sigma}_R(\omega',\mathbf{x}) e^{i\omega' t}\over \omega'-\omega} + {\Phi\over 2}\int d\omega' \tilde{\Sigma}_R(\omega',\mathbf{x})e^{i\omega't}\delta(\omega'-\omega)~.\nonumber\\
\end{eqnarray}
The first term corresponds, on the one hand, to the real part of the self-energy containing corrections to the scalar field mass, including the divergent zero-temperature piece that must be renormalized and contributes to the effective potential $V_{\rm eff}(\phi_c)$. The second term, on the other hand, includes the imaginary part of the self-energy, related to the field's decay in Fourier space via $\tilde{\Sigma}_R(\omega, \mathbf{p})=2i\omega \Gamma_\phi(\omega, \mathbf{p})$. Hence, the second term is simply $\Gamma_\phi i\omega\Phi e^{i\omega t}=\Gamma_\phi\dot\phi_c$, thus yielding the friction term in the Langevin-like equation, conventionally included in e.g. studies of reheating after inflation, as well as in \cite{Bastero-Gil:Initial_Conditions_Inflation, Rosa:2021gbe}:
\begin{eqnarray} \label{Langevin_non_local_2}
\partial^2\phi_c+ V_{\rm eff}'(\phi_c)+\Gamma_\phi\dot\phi_c=\xi~.
\end{eqnarray}

We note that in other dynamical settings one may also approximate the non-local dissipation term by a local friction term, for example in warm inflation where one may consider an adiabatic approximation, since the field is varying slowly on the typical timescale characterizing the dynamics of the heat bath. In these cases, however, the dissipation coefficient does not in general coincide with the perturbative decay width.

Finally, the KMS relation between the self-energy functions $\Sigma_F$ and $\Sigma_R$ given in Eq.~(\ref{KMS}) in momentum-space, valid when the heat bath fields are close to thermal equilibrium, yields the fluctuation-dissipation relation between the noise two-point correlation function and the dissipation coefficient (decay width) given in Eq.~(\ref{eq:thermal noise correlator}) at small energies/momenta, $\omega, p\lesssim T$.

\section{Evolution of the temperature during thermal inflation}\label{app:thermodynamical_considerations}

In our calculation we assumed that no significant entropy is produced during thermal inflation as a result of fluctuation-dissipation effects, i.e.~that $T \propto a^{-1}$.  In this appendix we aim to verify this assumption. The flaton field satisfies the Langevin-like equation \cite{Berera:Warm_Microphysical}: 
\begin{equation}\label{eq:langevin_equation_field_space}	
\ddot{\phi} + (3 H + \Gamma_\phi) \dot{\phi} - a^{-2} \nabla^2 \phi + m_{\text{eff}}^2 \phi = \xi~,
\end{equation}
and by multiplying both sides by $\dot\phi$ we obtain:
\begin{equation}
\dot{\rho}_\phi + 3 H (\rho_\phi + p_\phi) = \xi \dot{\phi} - \Gamma_\phi \dot{\phi}^2 + \alpha^2 T \dot{T} \phi^2 +  a^{-2} \partial_i (\dot{\phi} \partial_i \phi)~,
\end{equation}
where the field's energy density and pressure are given by:
\begin{equation}
	\begin{aligned}
		\rho_\phi &= \frac{1}{2}\dot{\phi}^2 + \frac{1}{2} a^{-2} \partial_i \phi \partial_i \phi + V(\phi)~,\\
		p_\phi &= \frac{1}{2}\dot{\phi}^2 - \frac{1}{6} a^{-2} \partial_i \phi \partial_i \phi - V(\phi)~.\\
	\end{aligned}
\end{equation}
Conservation of the full energy-momentum tensor then yields the following continuity equation for the radiation energy density:
\begin{equation}\label{eq:radiation_averaged}
	\dot{\rho}_R + 4 H \rho_R = - \braket{\xi \dot{\phi}} + \Gamma_\phi \braket{\dot{\phi}^2} -\alpha^2 T \dot{T} \braket{\phi^2} - a^{-2} \braket{\partial_i (\dot{\phi} \partial_i \phi})~.
\end{equation}
We note that the radiation energy density is an ensemble average over the energy density of the relativistic degrees of freedom, which justifies considering also the thermal average of the terms on the right-hand side of the above equation. Here we have also neglected the sub-leading corrections to the radiation energy and entropy densities from the fermions' finite mass, $\sim g\braket{\sqrt{\phi^2}}\sim gT\ll T$.

Using the field solutions we obtained for the correlators\footnote{$ \braket{\bm{\nabla} (\dot{\phi} \bm{\nabla} \phi}) = - \int \frac{d^3 k_1}{(2 \pi)^3} \frac{d^3 k_2}{(2 \pi)^3}\braket{ \dot{\phi}_{k_1} \phi_{k_2}} \bm{k}_2 \cdot ( \bm{k}_1 + \bm{k}_2) \exp{[i \bm{x} \cdot (\bm{k}_1 + \bm{k}_2 )]} = 0~,$ since the integral of this delta function is non-zero if and only if $\bm{k}_1 = - \bm{k}_2$.}:
\begin{equation}
	\begin{aligned}
		\braket{\xi \dot{\phi}} &= \frac{\pi}{6} \Gamma_\phi T^4~,\\
		\Gamma_\phi \braket{\dot{\phi}^2} &= \frac{\pi}{6} \Gamma_\phi T^4 (1 - \delta)~,\\
		\alpha^2 T \dot{T} \braket{\phi^2} &= \frac{ \alpha^2}{2 \pi} T^3 \dot{T}\bigg[ 1 - \frac{\alpha}{\pi} \arctan \bigg( \frac{\pi}{\alpha}\bigg)\bigg] (1 - \delta) \approx \frac{15 \alpha^2}{4 \pi^3 g_*}  (1 - \delta)\dot{\rho}_R~,\\
		\braket{\partial_i (\dot{\phi} \partial_i \phi} &= 0~.
	\end{aligned}
\end{equation}
Note that the third term is related to the time-dependence of the thermal flaton mass, and yields a contribution to the variation of the radiation energy density comparable to the above-mentioned sub-leading corrections from the fermions' non-vanishing mass. For consistency, we thus neglect this term, and obtain:
\begin{equation}
\dot{\rho}_R + 4H \rho_R = -{5/\pi\over g_*}\Gamma_\phi \rho_R \delta~.
\end{equation}
From this we immediately see that the right-hand side can only be significant if $\Gamma_\phi\gtrsim H$, but this implies a quick thermalization of the flaton field such that $\delta \to 0$ exponentially fast, thus making this term negligible. This simply reflects the balance between the effects of fluctuations and dissipation as the flaton field reaches an equilibrium with the radiation bath. Note, furthermore, that the term on the right-hand side is suppressed by the relative contribution of the flaton to the number of relativistic species in equilibrium, $(g_{*, \phi}-g_* )/ g_*$, as obtained in Section \ref{sec:curvature}. We therefore conclude that one may consistently assume $\rho_R\propto a^{-4}$ and hence that $T\propto a^{-1}$ during thermal inflation.

\section{Field correlation functions}\label{appendix:correlations}

Here we list the field correlation functions used to compute the curvature perturbation power spectrum. As we mentioned above, when integrating over momentum modes we consider a sharp cutoff at $k = \pi T_c$, which constitutes a good approximation to the behaviour of the noise correlation function \cite{Hiramatsu:Thermal_Fluctuations_Thermal_Inflation}. To compute the curvature perturbation power spectrum on super-horizon scales, $k\ll aH$, we consider the leading order results in $k / \alpha T_c \sim (k/aH)(M_0/M_P) a\ll 1$ considering $M_0< M_P$ and noting that in our convention $a<a_c=1$ above the critical temperature.

\subsection*{Mode correlators}

The building blocks of all field correlators are the equal time correlators between the field modes and their time derivatives. Writing $\phi_k$ and $\dot{\phi}_k$ in terms of the Green's function \eqref{eq:green_function_thermal_mass_domination}:
\begin{equation}
    \begin{aligned}
        \phi_k &= H^{-2} \int_{z_i}^{z} ds \; s^{-2} G_s(z, s) \xi_k(s)~,\\
        \dot{\phi}_k &= H^{-1} z  \int_{z_i}^{z} ds \; s^{-2} \partial_z G_s(z, s) \xi_k(s)~,
    \end{aligned}
\end{equation}
one finds:
\begin{equation}\label{eq: thermal_modes_correlators}
    \begin{aligned}
        \braket{\phi_k \phi_{k'}} &= (2 \pi)^3 \delta^3(\bm{k} + \bm{k}') \frac{T}{a^3 \omega_k^2} (1 - \delta)~,\\
        \braket{\dot{\phi}_k \dot{\phi}_{k'}} &= (2 \pi)^3 \delta^3 (\bm{k} + \bm{k}') \frac{T}{a^3} (1 - \delta)~,\\
        \braket{\phi_k \dot{\phi}_{k'}} &= - \frac{\Gamma_\phi}{2} \braket{\phi_k \phi_{k'}} = -(2 \pi)^3 \delta^3(\bm{k} + \bm{k}') \frac{T \Gamma_\phi}{2 a^3 \omega_k^2} (1 - \delta)~.
    \end{aligned}
\end{equation}

\subsection*{Field correlators}

To compute the total energy density \eqref{eq:total_energy_density_averaged} one needs to determine the field variance and the average kinetic and gradient energies. Expanding the field in terms of comoving momentum modes, these are given by:
\begin{equation}
    \begin{aligned}
		\braket{\phi^2} &= \int \frac{d^3 k}{(2 \pi)^3} \frac{d^3 k'}{(2 \pi)^3} \braket{\phi_{k} \phi_{k'}} \exp(i \bm{x} \cdot (\bm{k} + \bm{k}'))~,\\
        \braket{\dot{\phi}^2} &= \int \frac{d^3 k}{(2 \pi)^3} \frac{d^3 k'}{(2 \pi)^3} \braket{\dot{\phi}_{k} \dot{\phi}_{k'}} \exp(i \bm{x} \cdot (\bm{k} + \bm{k}'))~,\\
        \braket{\partial_i \phi \partial_i \phi} &= \int \frac{d^3 k}{(2 \pi)^3} \frac{d^3 k'}{(2 \pi)^3} \braket{\phi_{k} \phi_{k'}} \bm{k} \cdot \bm{k}' \exp(i \bm{x} \cdot (\bm{k} + \bm{k}'))~.
    \end{aligned}
\end{equation}
Inserting the mode correlation functions \eqref{eq: thermal_modes_correlators} and integrating over comoving  momenta up to $\pi T_c$ one obtains:
\begin{equation}
    \begin{aligned}
        \braket{\phi^2} &= \frac{T^2}{2 \pi} (1 - \delta) \bigg[1 -\frac{\alpha}{\pi} \arctan \bigg(\frac{\pi}{\alpha} \bigg) \bigg]~,\\
        \braket{\dot{\phi}^2} &= \frac{\pi T^4}{6} (1- \delta)~,\\
        \braket{\partial_i \phi \partial_i \phi} &=
		\frac{\pi}{2} a^2 T^4 (1 - \delta) \bigg[\frac{1}{3} - \bigg(\frac{\alpha}{\pi}\bigg)^2 + \bigg(\frac{\alpha}{\pi}\bigg)^3 \arctan \bigg(\frac{\pi}{\alpha} \bigg)\bigg]~.
    \end{aligned}
\end{equation}

\subsection*{Contributions to the power spectrum}

Consider the power spectrum of a generic correlator $\braket{X_i (0) X_j (\bm{x})}$, for example $X_1 = \phi$, $X_2 = \dot{\phi}$ and $X_3 = \partial_i \phi$, that appears in \eqref{eq:energy_density_perturbations_variance}:
\begin{equation}
    \int d^3 x \exp(-i \bm{k} \cdot \bm{x})\braket{X_i (0) X_j (\bm{x})}^2~.
\end{equation}
Note that, upon expanding each quantity $X_j(\mathbf{x})$ in terms of comoving momentum modes, this yields four momentum integrals and a volume integral. Two of the momentum integrals can be performed using the two delta functions appearing in the mode correlators \eqref{eq: thermal_modes_correlators}. Then, the volume integral will generate a delta function with the two surviving momentum modes:
\begin{equation}
    \int d^3 x \; \exp[- i \bm{x} \cdot (\bm{k}_1 + \bm{k}_2 + \bm{k})] =  (2 \pi)^3 \delta^3(\bm{k}_1 + \bm{k}_2 + \bm{k})~.
\end{equation}
After integrating this delta function over another of the 3-momentum variables, we are left with a single 3-dimensional integral over $\bm{k}$ that we need to compute in each case. In the following table we give the different contributions to the power spectrum in terms of their corresponding momentum integrals: 
\begin{table}[H]
        \centering
        \caption{Contributions to the power spectrum in Eq. \eqref{eq:density_perturbations}.}
		\begin{tabular}{ | l| l| l|}
			\hline
			field-field & $\frac{m_\text{eff}^4}{2}\int d^3 x \; \exp(-i \bm{k} \cdot \bm{x}) \braket{\phi(0) \phi(\bm{x})}^2$ & $(1 - \delta)^2  \frac{\alpha^3}{2(2 \pi)^3} \frac{T^5}{a^3} I_1 (k)$ \\ 
            \hline
			field-kinetic & $m_\text{eff}^2 \int d^3 x \; \exp(-i \bm{k} \cdot \bm{x}) \braket{\phi(0) \dot{\phi}(\bm{x})}^2$ & $ (1- \delta)^2 \frac{\alpha}{(2 \pi)^3} \big(\frac{3 \alpha^4}{32 \pi}\big)^2 \frac{T^5}{a^3} I_1 (k)$ \\
			\hline
            field-gradient & $a^{-2} m_\text{eff}^2 \int d^3 x \; \exp(-i \bm{k} \cdot \bm{x}) \braket{\phi(0) \partial_i\phi(\bm{x})}^2$ & $(1 - \delta)^2 \frac{\alpha^3}{(2 \pi)^3} \frac{T^5}{a^3} I_2(k)$\\
            \hline
            kinetic-kinetic & $\frac{1}{2} \int d^3 x \; \exp(-i \bm{k} \cdot \bm{x}) \braket{\dot{\phi}(0) \dot{\phi}(\bm{x})}^2$ & $(1 - \delta)^2  \frac{\alpha^3}{2(2 \pi)^3} \frac{T^5}{a^3} I_3 (k)$\\
            \hline
            kinetic-gradient & $a^{-2} \int d^3 x \; \exp(-i \bm{k} \cdot \bm{x}) \braket{\dot{\phi}(0) \partial_i\phi(\bm{x})}^2$ & $(1 - \delta)^2 \frac{\alpha}{(2 \pi)^3} \big(\frac{3 \alpha^4}{32 \pi}\big)^2 \frac{T^5}{a^3} I_2 (k)$\\
            \hline
            gradient-gradient & $\frac{a^{-4}}{2} \int d^3 x \; \exp(-i \bm{k} \cdot \bm{x}) \braket{\partial_i \phi(0) \partial_j \phi(\bm{x})}^2$ & $(1 - \delta)^2\frac{\alpha^3}{2(2 \pi)^3}  \frac{T^5}{a^{3}} I_4(k)$\\
            \hline
		\end{tabular}
	\end{table}
The momentum integrals can be expressed in terms of the normalized comoving momentum  $\bm{y}=\bm{k}/\alpha T_c$ with norm $0<y < \pi/\alpha$. These are given by:
\begin{equation}
    \begin{aligned}
        I_1(k) &= \int dy \, d\Omega \; \frac{y^2}{(y^2 + 1) [(\bm{y} + \bm{k}/( \alpha T_c))^2 + 1]}~,\\
        I_2(k) &= \int dy \, d \Omega \; \frac{ y^2 \bm{y} \cdot (\bm{y} + \bm{k}/(\alpha T_c))}{(y^2 + 1) [(\bm{y} + \bm{k}/( \alpha T_c))^2 + 1]}~,\\
        I_3(k) &= \int dy \, d \Omega \; y^2 = \frac{4 \pi^4}{3 \alpha^3}~,\\
        I_4 (k) &=\int dy \, d \Omega \; \frac{ y^2 [\bm{y} \cdot (\bm{y} + \bm{k}/(\alpha T_c))]^2}{(y^2 + 1) [(\bm{y} + \bm{k}/( \alpha T_c))^2 + 1]}~,
    \end{aligned}
\end{equation}
where $d\Omega$ denotes integration over the solid angle in momentum space. Except for $I_3(k)$, all integrals above depend non-trivially on $k$. To leading order in $k / \alpha T_c$ these integrals are given by:
\begin{equation}
    \begin{aligned}
        I_1(k) &\simeq 4 \pi \bigg[ - \frac{1}{2} \frac{\pi \alpha}{\alpha^2 + \pi^2} + \frac{1}{2}\arctan(\pi/\alpha)\bigg]~,\\
        I_2(k) &\simeq 4 \pi \bigg[ \frac{\pi}{\alpha} + \frac{1}{2} \frac{\alpha \pi}{\alpha^2 + \pi^2} - \frac{3}{2} \arctan(\pi/\alpha) \bigg]~,\\
        I_4 (k) &\simeq 4 \pi \bigg[ - \frac{2\pi}{\alpha} + \frac{1}{3} \frac{\pi^3}{\alpha^3} -  \frac{1}{2} \frac{\alpha \pi}{\alpha^2 + \pi^2} + \frac{5}{2} \arctan(\pi/\alpha) \bigg]~,
    \end{aligned}
\end{equation}
which are the expressions used to compute the curvature perturbation power spectrum \eqref{eq:density_perturbations} given in the main body of this article.

\bibliography{references}

\end{document}